# Bayesian influence diagnostics and outlier detection for meta-analysis of diagnostic test accuracy

Yuki Matsushima
Department of Statistical Science, School of Multidisciplinary Sciences, The Graduate University for Advanced Studies, Tokyo, Japan
Department of Biometrics, Otsuka Pharmaceutical Co Ltd, Tokyo, Japan
ORCID: http://orcid.org/0000-0002-6868-2699

Hisashi Noma, PhD[*]
Department of Data Science, The Institute of Statistical Mathematics, Tokyo, Japan
ORCID: http://orcid.org/0000-0002-2520-9949

Tomohide Yamada, MD, PhD
Department of Diabetes and Metabolic Diseases, Graduate School of Medicine, The University of Tokyo, Tokyo, Japan

Toshi A. Furukawa, MD, PhD
Departments of Health Promotion and Human Behavior and of Clinical Epidemiology, Kyoto University Graduate School of Medicine/School of Public Health, Kyoto, Japan

*Corresponding author: Hisashi Noma
Department of Data Science, The Institute of Statistical Mathematics
10-3 Midori-cho, Tachikawa, Tokyo 190-8562, Japan
TEL: +81-50-5533-8440
e-mail: noma@ism.ac.jp

**Summary**

Meta-analyses of diagnostic test accuracy (DTA) studies have been gaining prominence in research in clinical epidemiology and health technology development. In these DTA meta-analyses, some studies may have markedly different characteristics from the others, and potentially be inappropriate to include. The inclusion of these "outlying" studies might lead to biases, yielding misleading results. In addition, there might be influential studies that have notable impacts on the results. In this article, we propose Bayesian methods for detecting outlying studies and their influence diagnostics in DTA meta-analyses. Synthetic influence measures based on the bivariate hierarchical Bayesian random effects models are developed because the overall influences of individual studies should be simultaneously assessed by the two outcome variables and their correlation information. We propose four synthetic measures for influence analyses: (1) relative distance, (2) standardized residual, (3) Bayesian p-value, and (4) influence statistic on the area under the summary receiver operating characteristic curve. We also show conventional univariate Bayesian influential measures can be applied to the bivariate random effects models, which can be used as marginal influential measures. We illustrate the effectiveness of the proposed methods by applying them to a DTA meta-analysis of ultrasound in screening for vesicoureteral reflux among children with urinary tract infections.

# 1. Introduction

Evidence synthesis methods have been gaining prominence in diagnostic test accuracy (DTA) research in clinical epidemiology and health technology development [1,2]. Due to the methodological developments in DTA meta-analysis, the bivariate meta-analysis model is one of the standard methods for these studies, as it enables synthesis of the two primary correlated outcomes of diagnostic studies, sensitivity and false positive rate (FPR; = 1−specificity), thereby incorporating their correlation in statistical inference. In addition, the bivariate modeling framework provides a unified formulation that identifies the corresponding summary receiver operating characteristic (SROC) curve [3-7].

In DTA meta-analysis, there are systematic heterogeneities between studies in general, e.g., study designs, participant characteristics, regions, sites, cut-off of diagnostic markers, and outcome definitions. Therefore, heterogeneity of the diagnostic measure is common and random effects models are usually adopted [4,8]. However, some studies might have markedly different characteristics from others, and may exceed the degree of statistical heterogeneity that can be adequately explained by the random effects model. These "outlying studies" might lead to biases, potentially yielding misleading results. These biases might have serious influences on technology assessments and policy-making. Therefore, identification and influence diagnostics of the outlying studies are relevant to the practice of evidence synthesis research.

For outlier detection in DTA meta-analysis, several exploratory methods and graphical tools have been discussed in the literature [9-12], but they have limitations due to their heuristic and subjective approaches [13]. Recently, to address these issues, Negeri and Beyene [13] proposed more objective approaches based on the Reitsma's frequentist bivariate random effects model [4]. They proposed a bivariate residual-based diagnostic method and a test-based method using a mean-shift outlier model within the frequentist framework [13]. Besides, Bayesian influence diagnostic methods are another effective approaches for these problems [14]. In particular, the influence diagnostic methods of Carlin and Louis [14] have been widely applied to various statistical problems as useful tools for outlier detections. In evidence synthesis methods, Zhang et al. [15] recently developed detection and handling methods for trial-level outliers in network meta-analyses using the Carlin-Louis-type influence diagnostic methods. However, no such effective methods have been established for DTA meta-analyses, although the Bayesian hierarchical modeling is another effective approach for these meta-analyses [9,16].

In this article, we propose Bayesian methods for the identification and influential diagnostics of outlying studies in DTA meta-analyses. Especially, we develop synthetic influence measures based on the bivariate hierarchical Bayesian random effects models because the overall influences of individual studies should be simultaneously assessed by the two outcome variables and their correlation information. We propose four methods within the Bayesian framework of DTA meta-analyses to detect outlying or influential studies: (1) relative distance, (2) standardized residual, (3) Bayesian p-value, and (4) influence statistic on the area under the SROC curve. Note that outlying and influential studies are not the same, e.g., large studies can influential even if not outlying. Our methods aim to detect outliers and influential studies if they vary more than expected by chance. We illustrate the effectiveness of the proposed methods by applying them to a DTA meta-analysis of ultrasound in screening for vesicoureteral reflux among children with urinary tract infections [17].

The remainder of this paper is organized as follows. In Section 2, we briefly review the Bayesian bivariate model for DTA meta-analysis. In Section 3, we present the four methods for assessment of outlying studies and influence diagnostics. In Section 4, we demonstrate the effectiveness of these methods via application to a DTA meta-analysis of ultrasound in screening for vesicoureteral reflux among children with urinary tract infections [17]. Finally, we provide a discussion in Section 5.

**2. Bayesian bivariate hierarchical random effects model for DTA meta-analysis**

We consider the Bayesian bivariate hierarchical random effects model for sensitivity and FPR for meta-analysis of diagnostic test accuracy to address the between-studies heterogeneity [9,16]. Let $TP_i$, $FP_i$, $FN_i$ and $TN_i$ be the counts of true positive, false positive, false negative, and true negative participants in the $i$th study, respectively ($i = 1,2,\ldots,N$). Also, we denote the total numbers of positive and negative diagnoses in the $i$th study as $n_{Ai} = TP_i + FN_i, n_{Bi} = FP_i + TN_i$. Then, the logit-transformed sensitivity and FPR estimators are defined as $y_{Ai} = \text{logit}(TP_i/n_{Ai})$ and $y_{Bi} = \text{logit}(FP_i/n_{Bi})$.

Firstly, we consider the binomial probability model for $TP_i$ and $FP_i$,

$$TP_i \sim Bin(n_{Ai}, p_{Ai}), \quad FP_i \sim Bin(n_{Bi}, p_{Bi}).$$

where $p_{Ai}$ and $p_{Bi}$ are the sensitivity and FPR for the $i$th study, respectively. Then, we consider the random effects model for the logit-transformed binomial probability parameters, $\theta_{Ai} = \text{logit}(p_{Ai}), \theta_{Bi} = \text{logit}(p_{Bi})$,

$$\boldsymbol{\theta}_i \sim N(\boldsymbol{\mu}, \boldsymbol{\Sigma}), \quad \boldsymbol{\Sigma} = \begin{pmatrix} \sigma_A^2 & \rho\sigma_A\sigma_B \\ \rho\sigma_A\sigma_B & \sigma_B^2 \end{pmatrix},$$

where $\boldsymbol{\theta}_i = (\theta_{Ai}, \theta_{Bi})^T$ and $\boldsymbol{\mu} = (\mu_A, \mu_B)^T$, which is the summary logit-transformed sensitivity and FPR. $\sigma_A^2$ and $\sigma_B^2$ correspond to the heterogeneity variances of $\theta_{Ai}$ and $\theta_{Bi}$, and $\rho$ is their correlation coefficient. Note that $TP_i$ and $FP_i$ are assumed to be conditional independent given the random effect parameters $\theta_{Ai}$ and $\theta_{Bi}$. In Bayesian inference, the non-informative prior distributions are usually adopted, e.g., $\mu_A, \mu_B \sim N(0, 100), \sigma_A, \sigma_B \sim U(0, 10), \rho \sim U(-1, 1)$. Posterior inferences of the model parameters can be implemented using Markov Chain Monte Carlo (MCMC) [18,19].

## 3. Detections of outlying studies and influence diagnostics

*3.1 Relative distance*

To assess whether individual studies have remarkably different characteristics, we propose quantifiable measures that are suitable for DTA meta-analyses. The first measure is relative distance (RD), which was proposed by Zhang et al. [15] for network meta-analysis in order to assess the influence on the estimate of mean parameter by deleting the $i$th study from the calculation (i.e. leave-one-out cross-validation). This was proposed as a measure resembling Cook's distance for linear regression analyses [20]. The marginal RD for sensitivity and FPR of the $i$th study are defined by straightforwardly applying the RD measure of Zhang et al. [15],

$$RD_{Ai} = \left| \frac{\hat{\eta}_A - \hat{\eta}_{A(i)}}{\hat{\eta}_A} \right|, RD_{Bi} = \left| \frac{\hat{\eta}_B - \hat{\eta}_{B(i)}}{\hat{\eta}_B} \right|,$$

where $\hat{\eta}_A = \text{logit}^{-1}(\hat{\mu}_A)$, $\hat{\eta}_{A(i)} = \text{logit}^{-1}(\hat{\mu}_{A(i)})$, $\hat{\eta}_B = \text{logit}^{-1}(\hat{\mu}_B)$ and $\hat{\eta}_{B(i)} = \text{logit}^{-1}(\hat{\mu}_{B(i)})$. $\hat{\mu}_A$ and $\hat{\mu}_B$ are the estimators of $\mu_A$ and $\mu_B$ obtained from all the data, and $\hat{\mu}_{A(i)}$ and $\hat{\mu}_{B(i)}$ are those obtained from the leave-one-out data without the $i$th study. These estimators are obtained by the Bayesian hierarchical random effects model in Section 2 (usually, the posterior means are adopted). Studies with large RD are judged as influential in the sense of marginal measures. Note that the proposed measures are defined on the back-transformed scales, because similar measures on the logit-transformed scales cannot be defined when the denominators of the relative measures are zero, i.e., when the overall sensitivity or FPR is 50%. Therefore, we propose to adopt the back-transformed measures to circumvent the irregular property.

Besides, in DTA meta-analysis, the primary target to be estimated is substantially the grand mean parameter $\boldsymbol{\mu}$, and it is estimated using both the sensitivity and FPR information, as well as their correlation information. Thus, it is more reasonable that the influences are assessed as a two-dimensional measure. We then propose a synthetic measure for DTA meta-analysis, synthetic relative distance (SRD), which accounts for both sensitivity and FPR,

$$SRD_i = \frac{\sqrt{(\hat{\eta}_A - \hat{\eta}_{A(i)})^2 + (\hat{\eta}_B - \hat{\eta}_{B(i)})^2}}{\sqrt{\hat{\eta}_A^2 + \hat{\eta}_B^2}}.$$

Note that SRD is defined as a relative Euclidean distance in two-dimensional space that quantifies the divergence of the obtained estimates from the leave-one-out dataset. Another synthetic measure might be the mean of the two marginal measures of $RD_{Ai}$ and $RD_{Bi}$,

$$ARD_i = \frac{1}{2}\left(\left|\frac{\hat{\eta}_A - \hat{\eta}_{A(i)}}{\hat{\eta}_A}\right| + \left|\frac{\hat{\eta}_B - \hat{\eta}_{B(i)}}{\hat{\eta}_B}\right|\right).$$

which corresponds to the average RD (ARD) measure of Zhang et al. [15]. Geometrically, the ARD is solely an averaged relative measure of two one-dimensional Euclidean distances.

In addition, to assess the influences synthetically, another approach is to evaluate them using a synthetic diagnostic measure. A well-used measure is the diagnostic odds ratio (DOR) [21], which is defined as a ratio of positive likelihood ratio ($LR_{(+)}$) and negative likelihood ratio ($LR_{(-)}$),

$$DOR = \frac{LR_{(+)}}{LR_{(-)}} = \frac{p_{Ai}/p_{Bi}}{(1 - p_{Ai})/(1 - p_{Bi})}.$$

The influences on sensitivity and FPR can also be evaluated jointly by the relative distance using DOR,

$$RD_{DORi} = \left|\frac{\widehat{DOR} - \widehat{DOR}_{(i)}}{\widehat{DOR}}\right|.$$

Here $\widehat{DOR}$ is estimated from the full data by $\widehat{DOR} = \exp(\hat{\mu}_A - \hat{\mu}_B)$. $\widehat{DOR}_{(i)}$ is the estimate obtained from the full data, except for the $i$ th study, by $\widehat{DOR}_{(i)} = \exp(\hat{\mu}_{A(i)} - \hat{\mu}_{B(i)})$.

### 3.2 Standardized residual

In another approach based on leave-one-out cross validation, Carlin and Louis [14] and Zhang et al. [15] proposed using the standardized residual (SR). The SR is defined as a deviation measure, which is the difference in the observed outcome of the *i*th study and the posterior predictive mean obtained from the leave-one-out dataset, standardized by the posterior predictive standard deviation. Zhang et al. [15] proposed the SR as a univariate influence diagnostic measure for network meta-analysis; it can be straightforwardly applied to DTA analyses,

$$SR_{Ai} = \frac{y_{Ai} - E_{y_{Ai}|\mathbf{y}_{(i)}}(y_{Ai}|\mathbf{y}_{(i)})}{\sqrt{Var_{y_{Ai}|\mathbf{y}_{(i)}}(y_{Ai}|\mathbf{y}_{(i)})}}, SR_{Bi} = \frac{y_{Bi} - E_{y_{Bi}|\mathbf{y}_{(i)}}(y_{Bi}|\mathbf{y}_{(i)})}{\sqrt{Var_{y_{Bi}|\mathbf{y}_{(i)}}(y_{Bi}|\mathbf{y}_{(i)})}},$$

where $E_{y_{Ai}|\mathbf{y}_{(i)}}(y_{Ai}|\mathbf{y}_{(i)})$ and $E_{y_{Bi}|\mathbf{y}_{(i)}}(y_{Bi}|\mathbf{y}_{(i)})$ are the posterior predictive means, and $Var_{y_{Ai}|\mathbf{y}_{(i)}}(y_{Ai}|\mathbf{y}_{(i)})$ and $Var_{y_{Bi}|\mathbf{y}_{(i)}}(y_{Bi}|\mathbf{y}_{(i)})$ are the posterior predictive variances obtained from the leave-one-out dataset without the *i*th study. In addition, these measures only reflect marginal information. Similar to RD, we propose a synthetic measure called the synthetic standardized residual (SSR),

$$SSR_i = \left(\mathbf{y}_i - E_{\mathbf{y}_i|\mathbf{y}_{(i)}}(\mathbf{y}_i|\mathbf{y}_{(i)})\right)^T V_{\mathbf{y}_i|\mathbf{y}_{(i)}}(\mathbf{y}_i|\mathbf{y}_{(i)})^{-1} \left(\mathbf{y}_i - E_{\mathbf{y}_i|\mathbf{y}_{(i)}}(\mathbf{y}_i|\mathbf{y}_{(i)})\right)$$

where $\mathbf{y}_i = (y_{Ai}, y_{Bi})^T$, $E_{\mathbf{y}_i|\mathbf{y}_{(i)}}(\mathbf{y}_i|\mathbf{y}_{(i)}) = \left(E_{y_{Ai}|\mathbf{y}_{(i)}}(y_{Ai}|\mathbf{y}_{(i)}), E_{y_{Bi}|\mathbf{y}_{(i)}}(y_{Bi}|\mathbf{y}_{(i)})\right)^T$. $V_{\mathbf{y}_i|\mathbf{y}_{(i)}}(\mathbf{y}_i|\mathbf{y}_{(i)})$ is the posterior predictive covariance matrix obtained from the leave-one-out dataset lacking the *i*th study. SSR is derived as an extended measure of SR as a two-dimensional measure, whose denominator incorporates their variance–covariance information, and is given as quadratic form. In calculations of $E_{\mathbf{y}_i|\mathbf{y}_{(i)}}(\mathbf{y}_i|\mathbf{y}_{(i)})$ and

$V_{\mathbf{y}_i|\mathbf{y}_{(i)}}(\mathbf{y}_i|\mathbf{y}_{(i)})$, the following conditional predictive distribution [14] is used,

$$f(\mathbf{y}_i|\mathbf{y}_{(i)}) = \frac{f(\mathbf{y}_i)}{f(\mathbf{y}_{(i)})} = \int f(\mathbf{y}_i|\boldsymbol{\xi}, \mathbf{y}_{(i)}) p(\boldsymbol{\xi}|\mathbf{y}_{(i)}) \, d\boldsymbol{\xi}.$$

Here, $\boldsymbol{\xi}$ indicates an entire parameter vector, $\boldsymbol{\xi} = (\boldsymbol{\mu}^T, \sigma_A^2, \sigma_B^2, \rho)^T$. Note that the SSR is defined as the standardized deviation in two-dimensional space that quantifies the divergence of the observed statistics of the $i$th study from those of the leave-one-out dataset. In addition, the average SR (ASR), which was proposed by Zhang et al. [15], can also be applied to the DTA analyses as another synthetic measure,

$$ASR_i = \frac{1}{2}\left( \left| \frac{y_{Ai} - E_{y_{Ai}|\mathbf{y}_{(i)}}(y_{Ai}|\mathbf{y}_{(i)})}{\sqrt{Var_{y_{Ai}|\mathbf{y}_{(i)}}(y_{Ai}|\mathbf{y}_{(i)})}} \right| + \left| \frac{y_{Bi} - E_{y_{Bi}|\mathbf{y}_{(i)}}(y_{Bi}|\mathbf{y}_{(i)})}{\sqrt{Var_{y_{Bi}|\mathbf{y}_{(i)}}(y_{Bi}|\mathbf{y}_{(i)})}} \right| \right).$$

Also, we can discuss the standardized measure of DOR as a synthetic diagnostic measure,

$$SR_{DORi} = \frac{\log \widehat{DOR}_i - E_{\log DOR_i|\mathbf{y}_{(i)}}(\log DOR_i | \mathbf{y}_{(i)})}{\sqrt{Var_{\log DOR_i|\mathbf{y}_{(i)}}(\log DOR_i | \mathbf{y}_{(i)})}}.$$

Here, $\log \widehat{DOR}_i$ is the observed log-transformed diagnostic odds ratio for the $i$th study, and $E_{\log DOR_i|\mathbf{y}_{(i)}}(\log DOR_i | \mathbf{y}_{(i)})$ and $Var_{\log DOR_i|\mathbf{y}_{(i)}}(\log DOR_i | \mathbf{y}_{(i)})$ are the posterior predictive mean and variance of DOR obtained from the leave-one-out dataset, respectively. Because DOR is defined as an odds ratio of $p_{Ai}$ and $p_{Bi}$, the posterior predictive distribution can be straightforwardly constructed in the MCMC.

### *3.3 Bayesian p-value*

The Bayesian p-value is a well-established measure for posterior predictive model checking [14,22], which evaluates the discrepancy between the observed data and the posterior predictive samples obtained from the hierarchical Bayesian model by p-value.

To introduce the Bayesian p-value, we define the following discrepancy measures for sensitivity and FPR, respectively [14,22]:

$$D_i^A(y_{Ai}, \xi) = \frac{[y_{Ai} - E_{y_{Ai}|\xi}(y_{Ai}|\xi)]^2}{Var_{y_{Ai}|\xi}(y_{Ai}|\xi)}, D_i^B(y_{Bi}, \xi) = \frac{[y_{Bi} - E_{y_{Bi}|\xi}(y_{Bi}|\xi)]^2}{Var_{y_{Bi}|\xi}(y_{Bi}|\xi)}$$

where $E_{y_{Ai}|\xi}(y_{Ai}|\xi)$ and $E_{y_{Bi}|\xi}(y_{Bi}|\xi)$ are the posterior predictive means, and $Var_{y_{Ai}|\xi}(y_{Ai}|\xi)$ and $Var_{y_{Bi}|\xi}(y_{Bi}|\xi)$ are the posterior predictive variances. Then, we consider similar discrepancy measures defined by hypothetical future observations $y_{Ai}^*$ and $y_{Bi}^*$, $D_i^A(y_{Ai}^*, \xi)$ and $D_i^B(y_{Bi}^*, \xi)$, where $y_{Ai}^*$ and $y_{Bi}^*$ are posterior predictive samples generated by MCMC. The Bayesian p-values are defined as measures of extremeness of observed data from these measures, defined for sensitivity and FPR, respectively.

$$P_{D_i^A} = P[D_i^A(y_{Ai}^*, \xi) > D_i^A(y_{Ai}, \xi)|y] = \int P[D_i^A(y_{Ai}^*, \xi) > D_i^A(y_{Ai}, \xi)|\xi]p(\xi|y)\,d\xi,$$

$$P_{D_i^B} = P[D_i^B(y_{Bi}^*, \xi) > D_i^B(y_{Bi}, \xi)|y] = \int P[D_i^B(y_{Bi}^*, \xi) > D_i^B(y_{Bi}, \xi)|\xi]p(\xi|y)\,d\xi.$$

If the p-values $P_{D_i^A}$ or $P_{D_i^B}$ are small (e.g. < 0.15 or 0.10), it indicates that the observed data are extreme relative to the posterior predictive distribution. Then, the corresponding study is considered to be an outlying study, and possibly an influential one.

Although the previous two p-values are marginal diagnostic measures, we can also discuss another Bayesian p-value using the synthetic discrepancy measure defined in the previous sections. We use the two-dimensional discrepancy measure to define the diagnostic measure,

$$SD_i(\boldsymbol{y}_i, \xi) = \left(\boldsymbol{y}_i - E_{\boldsymbol{y}_i|\xi}(\boldsymbol{y}_i|\xi)\right)^T V_{\boldsymbol{y}_i|\xi}(\boldsymbol{y}_i|\xi)^{-1} \left(\boldsymbol{y}_i - E_{\boldsymbol{y}_i|\xi}(\boldsymbol{y}_i|\xi)\right),$$

and propose the Bayesian p-value using the two-dimensional information,

$$P_{SD_i} = P[SD_i(\boldsymbol{y}_i^*, \xi) > SD_i(\boldsymbol{y}_i, \xi)|\boldsymbol{y}] = \int P[SD_i(\boldsymbol{y}_i^*, \xi) > SD_i(\boldsymbol{y}_i, \xi)|\xi]p(\xi|\boldsymbol{y})\,d\xi.$$

The Bayesian p-value reflects the discrepancy information in two-dimensional space, and would be suitable for assessing the influences in DTA meta-analyses. Also, we can simply define the Bayesian p-value for averaged marginal discrepancy measures according to Zhang et al. [15],

$$AD_i(\boldsymbol{y}_i, \xi) = \frac{[y_{Ai} - E_{y_{Ai}|\xi}(y_{Ai}|\xi)]^2}{Var_{y_{Ai}|\xi}(y_{Ai}|\xi)} + \frac{[y_{Bi} - E_{y_{Bi}|\xi}(y_{Bi}|\xi)]^2}{Var_{y_{Bi}|\xi}(y_{Bi}|\xi)},$$

$$P_{AD_i} = P[AD_i(\boldsymbol{y}_i^*, \xi) > AD_i(\boldsymbol{y}_i, \xi)|\boldsymbol{y}] = \int P[AD_i(\boldsymbol{y}_i^*, \xi) > AD_i(\boldsymbol{y}_i, \xi)|\xi]p(\xi|\boldsymbol{y})\,d\xi.$$

We can also define the Bayesian p-value for DOR,

$$P_{D_{DORi}} = P[D(\log \widehat{DOR}_\iota^*, \xi) > D(\log \widehat{DOR}_\iota, \xi)|\boldsymbol{y}].$$

where

$$D_{DORi}(\log \widehat{DOR}_\iota, \xi) = \frac{[\log \widehat{DOR}_\iota - E_{\log DOR_i|\xi}(\log DOR_i|\xi)]^2}{Var_{\log DOR_i|\xi}(\log DOR_i|\xi)}.$$

Here, $\log \widehat{DOR}_\iota$ is the observed log-transformed diagnostic odds ratio for the $i$th study, and $\log \widehat{DOR}_\iota^*$ is hypothetical future data from a posterior predictive sample for the $i$th study. The posterior mean and variance for log-transformed DOR for the $i$th study are also obtained from MCMC.

*3.4 Influence on the area under the summary receiver operating characteristic curve*

In DTA meta-analyses, the AUC of the SROC curve is widely used to assess overall diagnostic ability. Influence analyses can be straightforwardly adapted to the AUC by deleting a study from the full data. Because the AUC can be directly interpreted as a summary measure of diagnostic performance, it would be useful to assess how influential each individual study is. After the posterior estimates of bivariate hierarchical random effects model in Section 2 are obtained, the SROC curve and AUC are estimated as follows [4].

1. Estimate the regression line of logit-transformed FPR on logit-transformed sensitivity,

$$E[Y_A|Y_B] = \hat{\mu}_A + \frac{\hat{\rho}\hat{\sigma}_A\hat{\sigma}_B}{\hat{\sigma}_B^2}(Y_B - \hat{\mu}_B).$$

2. Transform the regression line from logit scale back to the original (0–1) scale.
3. Estimate the AUC using the SROC curve by the ordinary method [2,5].

Then, influences of individual studies are evaluated using the following measure,

$$\Delta AUC_i = \widehat{AUC} - \widehat{AUC}_{(i)}$$

where $\widehat{AUC}$ is estimated from the full data and $\widehat{AUC}_{(i)}$ is the corresponding value after deletion of the $i$th study. A study with a large value of $\Delta AUC_i$ would be considered influential. Note that influence on AUC should be highly correlated with influence on DOR, as AUC and DOR are generally highly correlated.

## 4. Applications: DTA meta-analysis of ultrasound in screening for vesicoureteral reflux among children with urinary tract infections

To illustrate the proposed methods, we analyzed a dataset from a DTA meta-analysis of ultrasound in screening for vesicoureteral reflux (VUR) among children with urinary tract infections, performed by Shaikh et al. [17]. Because VUR, especially when severe, is linked to an increased risk of urinary tract infections and renal scarring, there is considerable interest in detecting it; however, voiding cystourethrography requires bladder catheterization and exposes children to radiation. Hence, alternative imaging tests for screening have been investigated. Shaikh et al. [17] conducted meta-analyses of two alternative imaging tests using DTA methods. We applied our methods to their meta-analysis, and in particular to their meta-analysis of the ultrasound in screening for VUR among children with urinary tract infections, which included 20 diagnostic studies. The data from these 20 studies are presented in Table 1 and the scatter plot of sensitivity and FPR is shown in Figure 1. There is a high degree of heterogeneity among studies, some of which could be explained by the use of pediatric urine collection bags and the inclusion of afebrile children[17]. In computations of these analyses, we used OpenBUGS version 3.2.3 [18,19] and R version 3.5.1 [23]. We conducted 120,000 iterations and discarded the first 20,000 iterations as the burn-in period for MCMC.

In Figure 2, we present the results for RD and SR. The vertical axis corresponds to the estimated values of RD or SR, and the horizontal axis corresponds to the study ID that was deleted from the dataset. The ten panels correspond to (a) RD for sensitivity, (b) RD for FPR, (c) SRD, (d) ARD, (e) RD for DOR, (f) SR for sensitivity, (g) SR for FPR, (h) SSR, (i) ASR, and (j) SR for DOR. For the synthetic RD, only studies 7 and 15 had SRD greater than 0.05, and this threshold value is equivalent to a 5% relative change from the original estimate; therefore, they would be considered as possible influential studies. Because study 15 had the highest sensitivity and FPR, and study 7 had the lowest sensitivity and a large number of subjects, they might be influential for estimating the pooled sensitivity and FPR. Studies 1, 7, 9, and 15 had an SSR larger than 4.61, which is in the upper 10th percentile of the chi-square distribution with two degrees of freedom. Therefore, we considered that these studies might be outlying. Studies 1, 9, and 15 had the highest or lowest FPR, and studies 7 and 15 had the highest or lowest sensitivity. In regard to DOR, studies 1, 7, 9, and 10 had RD > 0.05, and study 1 had the largest SR in absolute scale; therefore, they were considered as outlying studies and may have been influential on the DOR.

In Table 2, we show the Bayesian p-values for sensitivity, FPR, synthetic discrepancy measure, average discrepancy measure, and DOR. Studies 1 and 9 had the smallest p-values (< 0.15) for FPR, synthetic discrepancy measure, average discrepancy measure, and DOR; some of these values were < 0.05. Thus, studies 1 and 9 were identified as a potentially outlying studies based on the Bayesian p-values.

In Figure 3, we present the results of influence analyses on AUC, as determined by deleting each study individually. Although $\Delta AUC_i$ was smaller than 0.02 in absolute scale for most of the 20 studies, the values for studies 1 and 15 were -0.036 and 0.028, respectively. These two studies had relatively large $\Delta AUC_i$ values, suggesting that they are influential. Study 15 had the highest sensitivity and the highest FPR, and study 1 had the high FPR and the relatively low sensitivity; consequently, these studies had large influences on the AUC.

Finally, we compared the results of a DTA meta-analysis involving all 20 studies with those of a meta-analysis lacking the potentially outlying studies. In Table 3, we present the summaries of the potentially outlying studies, identified by the preceding analyses. Study 1 was identified as potentially outlying by four methods, and studies 7, 9, and 15 were identified by three methods. On the other hand, studies other than 1, 7, 9, 10, and 15 were not judged as outlying by any methods. Study 1 was published in 1986, which was the oldest study, and had the smallest number of subject (N = 29). The year of publication and the large chance variation due to small sample size might explain why study 1 was an outlier. After deleting these potentially outlying studies, the pooled sensitivity estimates ranged between 0.41 and 0.44, the pooled FPR estimates between 0.18 and 0.23, the areas under SROC curve between 0.565 and 0.634, and the pooled DOR estimates between 2.72 and 3.19. In Figure 4, we present plots of the SROC curves for these analyses. Some of the SROC curves did not differ significantly from the original result obtained using all 20 studies (the upper left panel). However, deleting study 1 or 15 seemed to change the shape of the SROC curve. Although potentially outlying studies have some impact on the estimate of AUC and the shape of the SROC curve, through detailed evaluations by our proposed methods, significant influences that changed the conclusions of this meta-analysis would not be detected. The conclusions of the original analysis are supported by these various influence analyses.

## 5. Discussion

In this paper, we proposed Bayesian methods for detecting outlying studies and influence diagnosis for DTA meta-analyses. Similar methods have been already discussed for network meta-analysis [15], but the marked characteristic of DTA meta-analysis is that the outcomes are evaluated by bivariate joint models, and the primary outputs are special diagnostic measures, e.g., DOR, the SROC curve, and its AUC [24]. Thus, special methods that are suitable for these assessments are needed for more appropriate scientific evaluations of diagnostic tools or markers.

In our methods, we need to specify some threshold values for explicitly identifying potential outlying studies. This might make it difficult to apply these methods in practice, but this is a common and general problem in outlier detections and influence diagnostics [25,15]. Adequacy depends on the case-by-case situation, and strict guidelines would be difficult to set up. However, several characteristics of these methods would be useful for these evaluations. First, if sensitivity or FPR is near 0, the denominator of RD $\hat{\eta}_A$ or $\hat{\eta}_B$ has a value near zero, and RD tends to have large value. On the other hand, if $\hat{\eta}_A$ or $\hat{\eta}_B$ has a relatively large absolute value, RD tends to be small even if there is a meaningful impact of deleting a study. Second, we consider the Bayesian bivariate hierarchical random effects model that is considered as a standard approach for meta-analysis of DTA; however, the use of logit transformation in the model might overemphasize small differences in sensitivity or FPR at the extremes. This is particularly an issue for SR or Bayesian p-value. Third, if the number of studies in a meta-analysis is small, the weight of each study becomes large, and detection methods based on leave-one-out cross-validation (i.e. RD, SR and influence on the AUC) tend to generate relatively large values. In addition, we should be careful about the multiplicity of these analyses. Because we must conduct a large number of analyses for these evaluations, extreme results may be observed by chance. This is also a common problem in influence diagnostics, and should be taken into account in practical situations.

Most of our proposed methods are synthetic measures that assess the magnitudes of influences on the two-dimensional space; however, in diagnostic analysis, the direction in ROC space is also critical. For example, if a study has extreme sensitivity but average FPR it might indicate outlying diagnostic accuracy. On the other hand, a study for which both sensitivity and FPR are 5% above the average might just indicate slight threshold heterogeneity. Under these situations, marginal measures for sensitivity or FPR may be effective supportive tools for evaluating the marginal properties (such as direction). These synthetic and marginal measures would both provide useful information.

Because our proposed measures assess the influence of studies from different criteria, the ability to detect outlying studies varies on a case-by-case basis. For example, RD and influence on the AUC will detect influential studies, and SR and Bayesian p-value will detect the outlying studies. Some measures assess the influences on sensitivity, FPR, or their combination, whereas others assess the influence on DOR or AUC. Therefore, no single measure is always versatile. All measures would provide useful information in some cases, and the results could be assessed from various viewpoints.

In our applications, we examined the results of a DTA meta-analysis lacking potentially outlying studies in order to evaluate the robustness of the original result. However, influential studies should not be removed naively from the main analysis. On the other hand, studies identified as outlying should be examined in term of their study designs, participant characteristics, regions, sites, cut-off of diagnostic markers, and outcome definitions.

In Section 4, we illustrated the effectiveness of the proposed methods by applying them to a DTA meta-analysis for ultrasound in screening for VUR among children with urinary tract infections. Our methods enabled us to identify potential outlying studies and quantitatively evaluate their influence on the overall results. Although any significant numerical evidence that there were some influential studies possibly to change the overall conclusions was not found, various quantitative evidence certainly supports the robustness of the main conclusions of that study. Such quantitative evidence would also be useful for policy-making and health technology assessments that use integrated evidence from these meta-analyses.

**Highlights**

- The evidence obtained from DTA meta-analyses has been widely applied for public health, clinical practices, health technology assessments, and policy making. The inclusion of "outlying" studies might lead to biases, yielding misleading evidence in DTA meta-analyses.

- We proposed new Bayesian methods for detecting outlying studies and their influence diagnostics in DTA meta-analyses. Especially, synthetic influence measures based on the bivariate hierarchical Bayesian random effects models are proposed because the overall influences of individual studies should be simultaneously assessed by the two outcome variables and their correlation information. The synthetic measures for influence analyses are (1) relative distance, (2) standardized residual, (3) Bayesian p-value, and (4) influence statistic on the area under the summary receiver operating characteristic curve. These methods enable effective evaluations of the outliers and their influences, and would provide precise knowledge in these evidence synthesis researches.

## Data Availability Statement

The DTA meta-analysis dataset used in Section 4 is part of the published data from Shaikh et al. [17].

## Acknowledgements

This study was supported by Grants-in-Aid for Scientific Research from the Japan Society for the Promotion of Science (Grant numbers: JP19H04074, JP17K19808).

**Table 1.** Results from 20 diagnostic accuracy studies of ultrasound in screening for VUR among children with urinary tract infections[†].

| ID | Study | TP | FP | FN | TN | Sensitivity [95% CI] | FPR [95% CI] |
|---|---|---|---|---|---|---|---|
| 1 | Alon (1986) | 7 | 9 | 11 | 2 | 0.39 [0.17, 0.64] | 0.82 [0.48, 0.98] |
| 2 | Alon (1999) | 8 | 2 | 30 | 60 | 0.21 [0.10, 0.37] | 0.03 [0.00, 0.11] |
| 3 | Calisti (2005) | 26 | 31 | 19 | 71 | 0.58 [0.42, 0.72] | 0.30 [0.22, 0.40] |
| 4 | Cleper (2004) | 6 | 7 | 7 | 44 | 0.46 [0.19, 0.75] | 0.14 [0.06, 0.26] |
| 5 | El Shenoufy (2009) | 5 | 2 | 8 | 30 | 0.38 [0.14, 0.68] | 0.06 [0.01, 0.21] |
| 6 | Goldman (2000) | 8 | 4 | 13 | 19 | 0.38 [0.18, 0.62] | 0.17 [0.05, 0.39] |
| 7 | Hoberman (2003) | 14 | 21 | 101 | 165 | 0.12 [0.07, 0.20] | 0.11 [0.07, 0.17] |
| 8 | Ismaili (2011) | 18 | 21 | 35 | 135 | 0.34 [0.22, 0.48] | 0.13 [0.09, 0.20] |
| 9 | Kim (2006) | 9 | 0 | 6 | 37 | 0.60 [0.32, 0.84] | 0.00 [0.00, 0.09] |
| 10 | Lee (2009) | 47 | 44 | 20 | 109 | 0.70 [0.58, 0.81] | 0.29 [0.22, 0.37] |
| 11 | Lee (2012) | 72 | 141 | 49 | 356 | 0.60 [0.50, 0.68] | 0.28 [0.24, 0.33] |
| 12 | Lopez Sastre (2007) | 25 | 55 | 29 | 118 | 0.46 [0.33, 0.60] | 0.32 [0.25, 0.39] |
| 13 | Mahant (2002) | 14 | 30 | 21 | 97 | 0.40 [0.24, 0.58] | 0.24 [0.17, 0.32] |
| 14 | Montini (2009) | 18 | 20 | 48 | 214 | 0.27 [0.17, 0.40] | 0.09 [0.05, 0.13] |
| 15 | Morin (1999) | 20 | 41 | 2 | 7 | 0.91 [0.71, 0.99] | 0.85 [0.72, 0.94] |
| 16 | Muller (2009) | 29 | 91 | 23 | 147 | 0.56 [0.41, 0.70] | 0.38 [0.32, 0.45] |
| 17 | Sheu (2013) | 60 | 70 | 93 | 250 | 0.39 [0.31, 0.47] | 0.22 [0.17, 0.27] |
| 18 | Soylu (2007) | 5 | 4 | 33 | 46 | 0.13 [0.04, 0.28] | 0.08 [0.02, 0.19] |
| 19 | Supavekin (2013) | 9 | 14 | 8 | 36 | 0.53 [0.28, 0.77] | 0.28 [0.16, 0.42] |
| 20 | Tsai (2012) | 40 | 96 | 20 | 64 | 0.67 [0.53, 0.78] | 0.60 [0.52, 0.68] |

[†] TP: true positive, FP: false positive, FN: false negative, TN: true negative, FPR: false positive rate, CI: confidence interval.

**Table 2.** Bayesian p-values.

| ID | Study | Sensitivity | FPR | Synthetic | Average | DOR |
|----|-------|-------------|-----|-----------|---------|-----|
| 1 | Alon (1986) | 0.2916 | 0.0973 | 0.1263 | 0.1469 | 0.0415 |
| 2 | Alon (1999) | 0.9777 | 0.5548 | 0.8394 | 0.8388 | 0.5940 |
| 3 | Calisti (2005) | 0.8087 | 0.9840 | 0.9713 | 0.9713 | 0.8500 |
| 4 | Cleper (2004) | 0.6655 | 0.8979 | 0.8898 | 0.9017 | 0.6457 |
| 5 | El Shenoufy (2009) | 0.6640 | 0.5759 | 0.7194 | 0.7750 | 0.4150 |
| 6 | Goldman (2000) | 0.9668 | 0.9404 | 0.9957 | 0.9967 | 0.9231 |
| 7 | Hoberman (2003) | 0.3992 | 0.7125 | 0.6521 | 0.6575 | 0.3802 |
| 8 | Ismaili (2011) | 0.9707 | 0.9922 | 0.9992 | 0.9992 | 0.9727 |
| 9 | Kim (2006) | 0.1565 | 0.0228 | 0.0244 | 0.0288 | 0.0012 |
| 10 | Lee (2009) | 0.5452 | 0.8798 | 0.8188 | 0.8212 | 0.5566 |
| 11 | Lee (2012) | 0.8247 | 0.9683 | 0.9760 | 0.9761 | 0.8310 |
| 12 | Lopez Sastre (2007) | 0.8894 | 0.8950 | 0.9803 | 0.9814 | 0.8441 |
| 13 | Mahant (2002) | 0.8790 | 0.8990 | 0.9794 | 0.9808 | 0.8373 |
| 14 | Montini (2009) | 0.9641 | 0.9614 | 0.9980 | 0.9977 | 1.0000 |
| 15 | Morin (1999) | 0.3340 | 0.7052 | 0.6092 | 0.5808 | 0.5474 |
| 16 | Muller (2009) | 0.9914 | 0.9347 | 0.9968 | 0.9967 | 0.9770 |
| 17 | Sheu (2013) | 0.9459 | 0.9311 | 0.9939 | 0.9941 | 0.9126 |
| 18 | Soylu (2007) | 0.3935 | 0.7932 | 0.6298 | 0.6659 | 0.4081 |
| 19 | Supavekin (2013) | 0.8332 | 0.9390 | 0.9780 | 0.9760 | 0.8979 |
| 20 | Tsai (2012) | 0.9816 | 0.8696 | 0.9852 | 0.9854 | 0.9138 |

**Table 3.** Comparison of pooled sensitivity and false positive rate, area under the SROC curve, and diagnostic odds ratio between the results of meta-analyses of all 20 diagnostic accuracy studies and the results obtained after deleting potential outlying studies [†].

|  | Outlying studies [‡] | Sensitivity [95% CI] | FPR [95% CI] | AUC [95% CI] | DOR [95% CI] |
|---|---|---|---|---|---|
| All studies | — | 0.44 [0.33, 0.56] | 0.22 [0.13, 0.34] | 0.588 [0.474, 0.676] | 2.82 [1.75, 4.60] |
| Without outlying or influential studies identified by | | | | | |
|   Relative distance | Study 7, 15 | 0.44 [0.35, 0.53] | 0.20 [0.12, 0.30] | 0.565 [0.441, 0.659] | 3.19 [1.98, 5.37] |
|   Standardized residual | Study 1, 7, 9, 15 | 0.43 [0.33, 0.53] | 0.20 [0.13, 0.29] | 0.625 [0.543, 0.687] | 3.04 [2.21, 4.28] |
|   Bayesian p-value | Study 1, 9 | 0.44 [0.32, 0.57] | 0.23 [0.14, 0.34] | 0.634 [0.560, 0.693] | 2.72 [1.98, 3.80] |
|   Diagnostic odds ratio | Study 1, 7, 9, 10 | 0.44 [0.33, 0.57] | 0.23 [0.13, 0.36] | 0.618 [0.557, 0.670] | 2.79 [2.00, 3.95] |
|   Influence analysis on the AUC | Study 1, 15 | 0.41 [0.31, 0.52] | 0.18 [0.12, 0.26] | 0.606 [0.471, 0.695] | 3.13 [2.12, 4.86] |

[†] SROC: summary receiver operating characteristic, FPR: false positive rate, CI: confidence interval, AUC: area under the summary receiver operating characteristic curve, DOR: diagnostic odds ratio.

[‡] Thresholds for identifying outlying or influential studies are 0.05 for relative distance, 4.61 (upper 10 percentile of $\chi_2^2$ distribution) for standardized residual, 0.15 for Bayesian p-value, and 0.02 for influence analysis on the AUC.

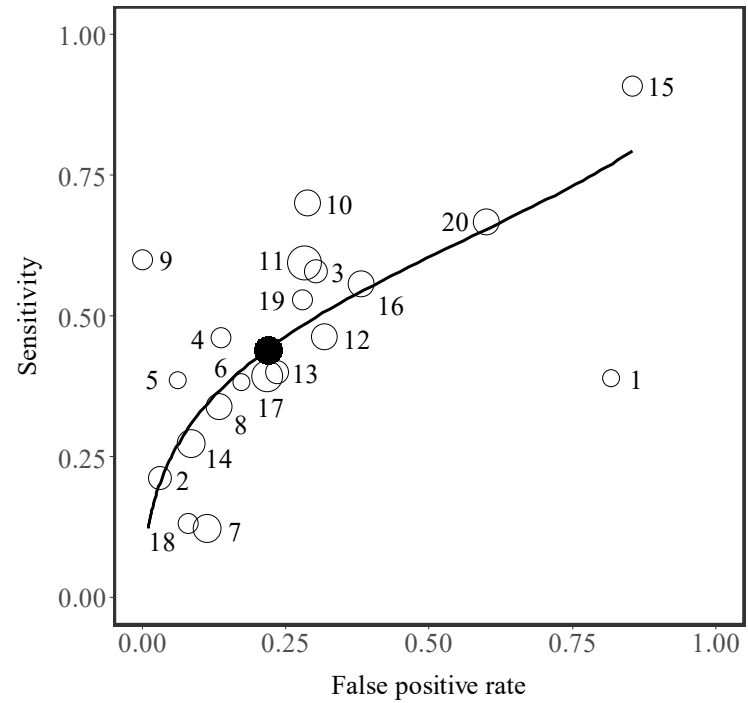

**Figure 1.** Scatter plot of sensitivity and false positive rate from 20 diagnostic accuracy studies of ultrasound in screening for VUR among children with urinary tract infections (white circles: sensitivity and false positive rate of individual study; black circle: pooled sensitivity and false positive rate).

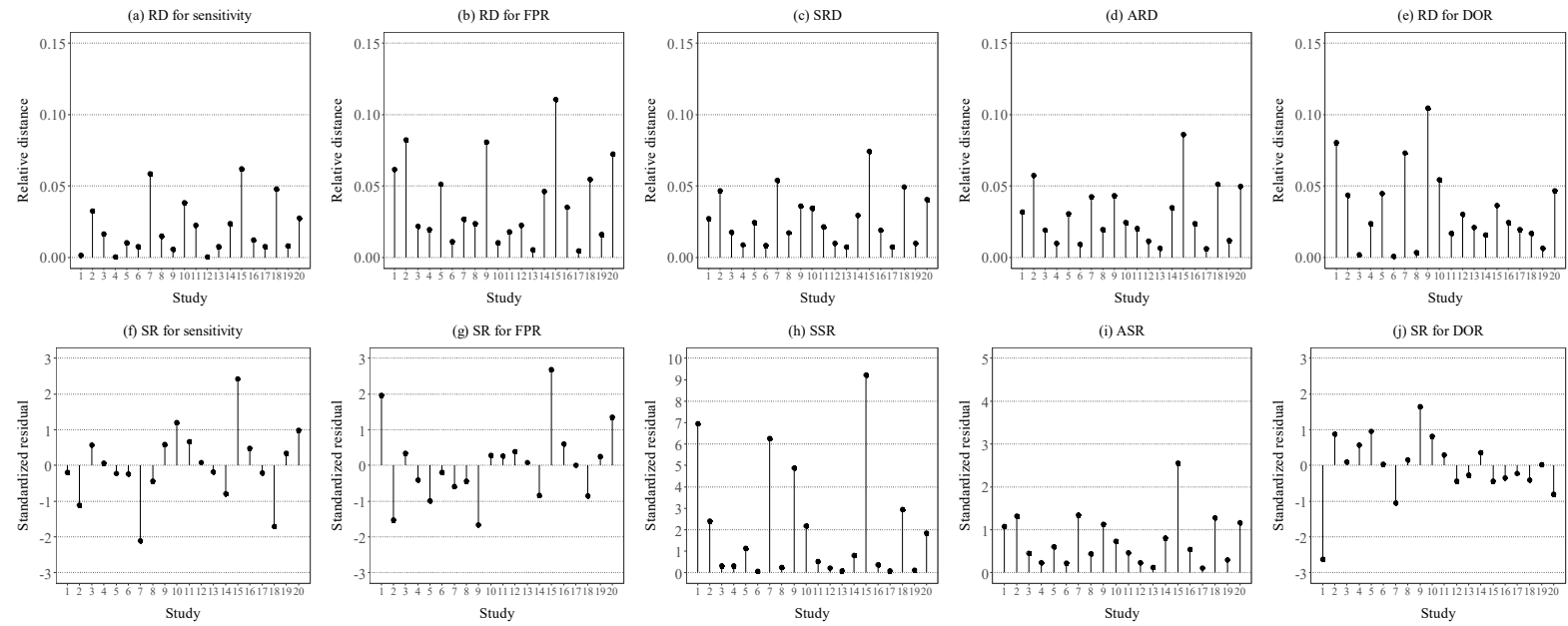

**Figure 2.** Relative distances and standardized residuals by leave-one-out influence analyses (RD: relative distance, SRD: synthetic relative distance, ARD: average relative distance, DOR: diagnostic odds ratio, SR: standardized residual, SSR: synthetic standardized residual, ASR: average standardized residual).

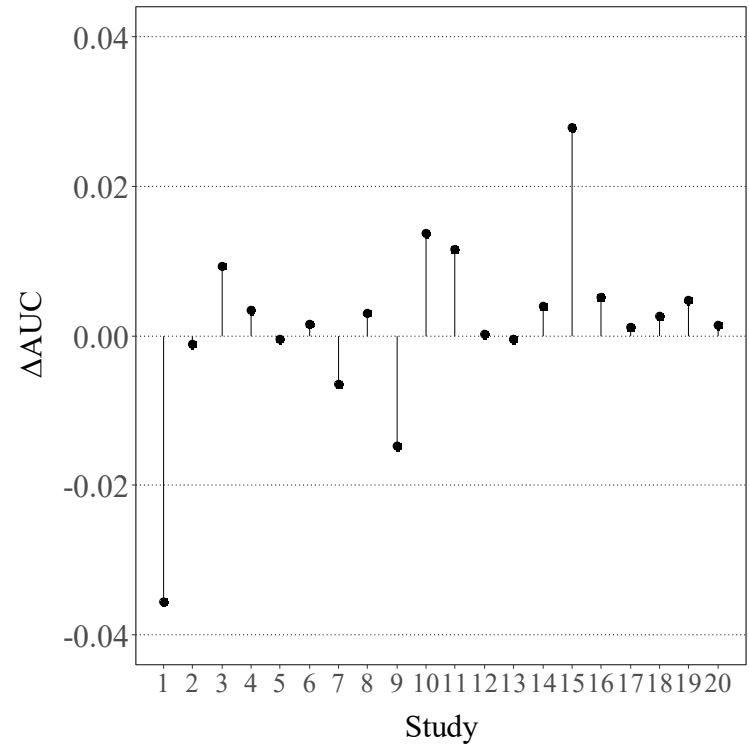

**Figure 3.** Results of influence analyses on the AUC.

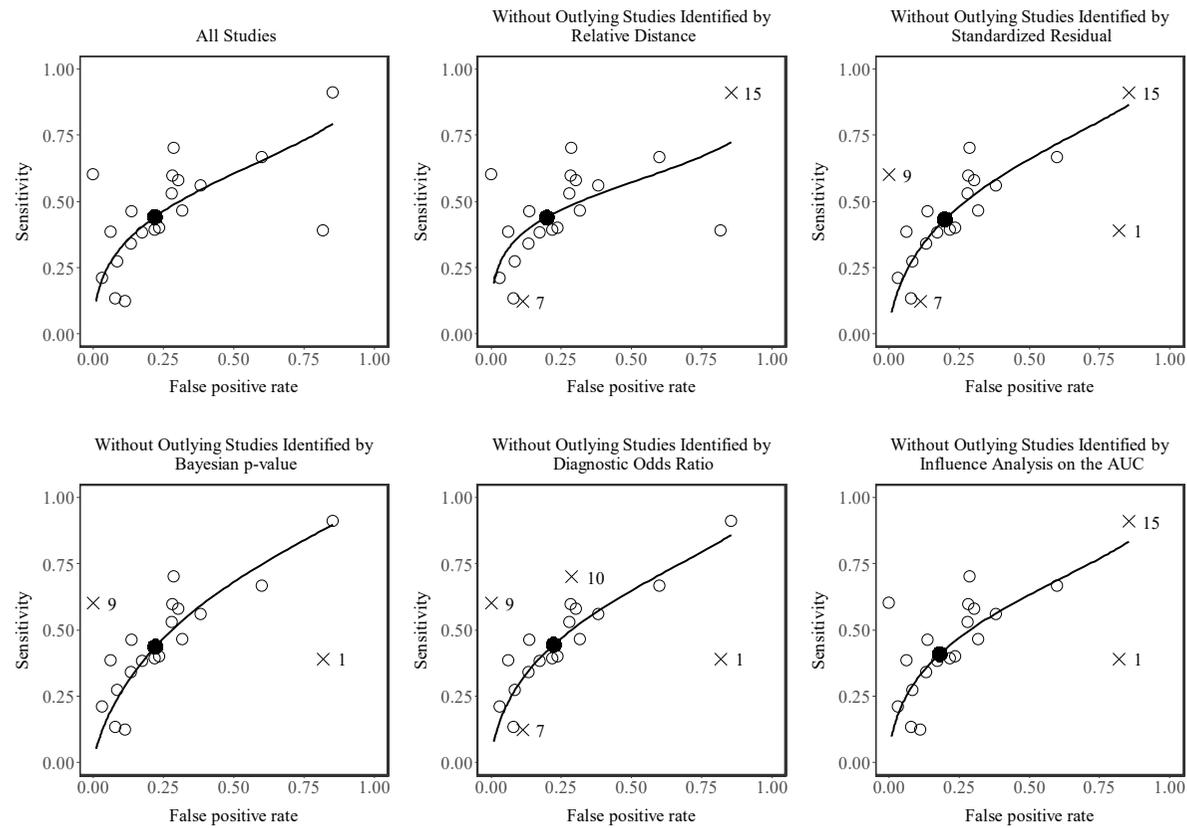

**Figure 4.** Scatter plot of sensitivity and false positive rate with and without potential outlying or influential studies (x: sensitivity and false positive rate of outlying or influential studies; white circles: sensitivity and false positive rate of other individual studies; black circle: pooled sensitivity and false positive rate). Thresholds for identifying outlying or influential studies are 0.05 for relative distance, 4.61 (upper 10 percentile of $\chi^2_2$ distribution) for standardized residual, 0.15 for Bayesian p-value, and 0.02 for influence analysis on the AUC. Outlying or influential studies identified by each method are as follows: studies 7 and 15 for relative distance; studies 1, 7, 9, and 15 for standardized residual; studies 1 and 9 for Bayesian p-value; studies 1, 7, 9, and 10 for diagnostic odds ratio; and studies 1 and 15 for influence analysis on the AUC.